\documentclass[twocolumn]{webofc}

\usepackage[varg]{txfonts}   
\usepackage{hyperref}
\usepackage{url}
\usepackage{mathrsfs}
\hypersetup{colorlinks=true,citecolor=blue,urlcolor=blue,linkcolor=blue}
\begin{document}

\title{ALICE FoCal overview}

\author{\firstname{Jonghan} \lastname{Park}\inst{1}\fnsep\thanks{\email{jonghan@cern.ch}} for the ALICE Collaboration}

\institute{University of Tsukuba}

\abstract{
The Forward Calorimeter (FoCal) is a new sub-detector in ALICE to be installed during the LHC Long Shutdown 3 for LHC Run 4. It consists of a highly-granular Si+W electromagnetic calorimeter combined with a conventional metal-scintillator hadronic calorimeter, covering a pseudorapidity interval of $3.2<\eta<5.8$. The FoCal is optimised to measure various physics quantities in the forward region, allowing exploration of the gluon density in hadronic matter down to $x\sim10^{-6}$, thus providing insights into non-linear QCD evolution at the LHC. These proceedings introduce the FoCal physics program and its corresponding performance. Additionally, the performance of the FoCal prototype will be presented.
}

\maketitle

\section{Introduction}
\label{intro}
The main scientific objective of the ALICE Forward Calorimeter (FoCal) is to study gluon saturation. In hadronic matter, the gluon density increases linearly with decreasing momentum fraction $x$, following a power law $xg(x) \sim x^{-\delta/2}$, where the exponent $\delta$ is determined by data fitting. However, at sufficiently small values of $x$, the gluon density becomes so high that partons start to overlap and recombine, causing a reduction in gluon density relative to linear projections~\cite{Armesto_2006}. Consequently, the growth of the gluon density cannot continue indefinitely as $x$ decreases; instead, the gluon density eventually saturates due to gluon self-interaction, forming a new state of gluon-saturated matter. This state is characterised by a saturation momentum $Q_{\rm sat}$; gluons with momentum below $Q_{\rm sat}$ experience saturation effects, while those with higher momentum follow linear QCD evolution. The saturation effects are more pronounced in heavy nuclei, as $Q_{\rm sat}^{2} \sim A^{1/3}$, where $A$ represents the nuclear mass~\cite{Gelis_2010}.

The FoCal will explore these novel phenomena through forward measurements of multiple electromagnetic and hadronic observables in hadronic pp and p--Pb collisions, as well as in ultra-peripheral p--Pb and Pb--Pb collisions. The data provided by FoCal will probe the partonic structure of hadronic matter and the nature of QCD evolution, reaching unprecedented momentum fraction down to $x \sim 10^{-6}$ for small momentum transfer $Q^{2}\approx4,{\rm GeV}/c$~\cite{loi,ALICE:2023fov}. FoCal will be installed during the LHC Long Shutdown 3 for data-taking in Run 4 (2029--2032). It will be positioned at $7\,{\rm m}$ from the interaction point of the ALICE detector, close to the beam pipe, enabling coverage of a very forward rapidity range, $3.2<\eta<5.8$. FoCal consists of two parts: the electromagnetic calorimeter (FoCal-E), optimised for measuring direct photons, neutral and vector mesons, and the hadronic calorimeter (FoCal-H), designed for photon isolation and jet measurements. These proceedings report on the FoCal physics program and its performance. Additionally, the performance of the FoCal detector prototype will be presented.

\section{FoCal physics performance}
The FoCal physics performance is studied through simulated pp, p--Pb and Pb--Pb collision events, utilising PYTHIA8~\cite{pythia8} and HIJING~\cite{hijing}, along with an idealised FoCal geometry implemented in GEANT3~\cite{geant}. Additionally, the performance for measurements in Ultra-Peripheral Collisions (UPCs) is assessed using STARlight simulations~\cite{starlight}. These performance projections assume that the integrated luminosity $\mathscr{L}_{\rm int}$ delivered during Run 4 will be $100\,{\rm pb^{-1}}$ for pp collisions at $\sqrt{s}=14\,{\rm TeV}$, $300\,{\rm nb^{-1}}$ for p--Pb collisions at $\sqrt{s}=8.8\,{\rm TeV}$ split into p--Pb and Pb--p configurations, and $7\,{\rm nb^{-1}}$ for Pb--Pb collisions at $\sqrt{s}=5.5\,{\rm TeV}$. This section will present several observables sensitive to the gluon saturation effect.\\[2mm]
\emph{- Direct photon measurements}

Direct photons are primarily produced at the parton interaction vertex via the Compton process $qg\rightarrow\gamma q$ at the LHC, providing access to the gluon density because they directly couple to incoming quarks and are not affected by final state effects. One of the key challenges in this measurement is distinguishing signal photons originating directly from the hard scattering process from decay photons. Enhancing the contribution of direct photons in FoCal can be achieved through three techniques: 1) isolation energy within a cone of a specified radius around the photon candidate in FoCal, 2) invariant mass of cluster pairs to reject photons from $\pi^{0}$ mesons ($\pi^{0}\rightarrow\gamma\gamma$),  and 3) the long axis distribution of the shower shape ellipse originating from decay photons with small opening angles. Using these techniques, the signal fraction can be increased up to 72\% at $p_{\rm T}=14\,{\rm GeV}/c$ (by a factor approximately 11)~\cite{focalphysicsperformance}. The physics impact of direct photon measurements is assessed by constructing FoCal pseudo-data based on existing PDFs. Statistical and systematic uncertainties for the data were estimated from NLO calculations of production cross section.

\begin{figure}[!hpt]
    \centering
    \includegraphics[width=0.48\textwidth]{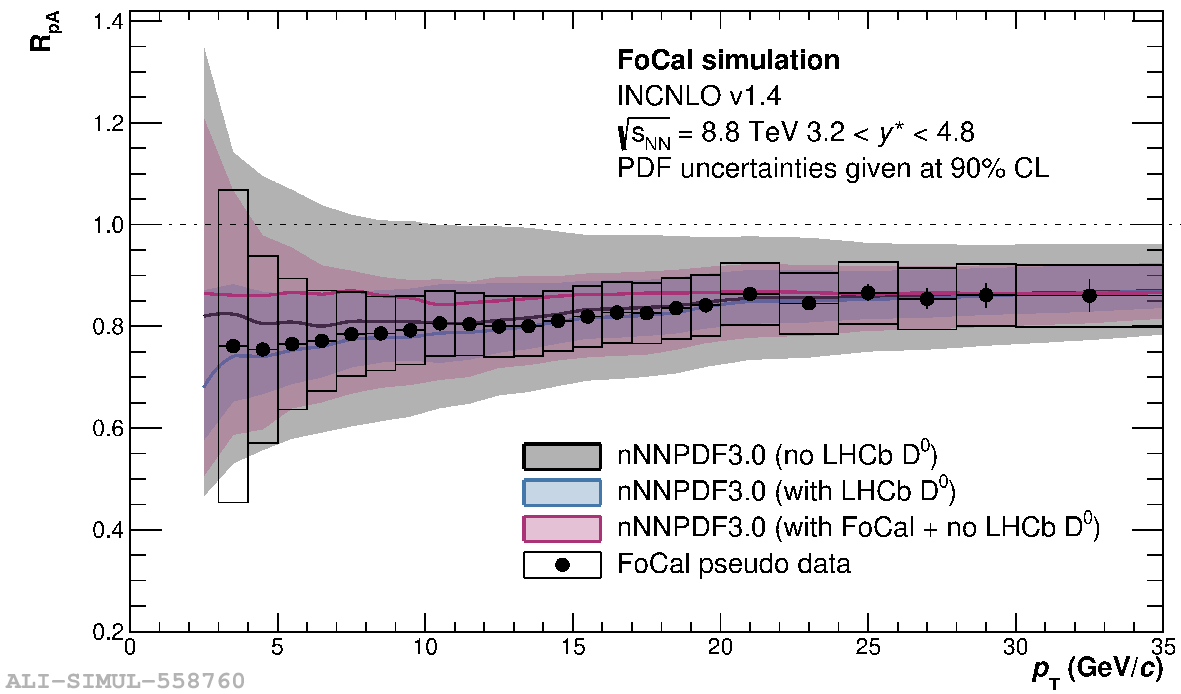}
    \caption{$R_{\rm pPb}$ for inclusive direct photons at $\sqrt{s_{\rm NN}}=8.8\,{\rm TeV}$ for FoCal pseudo-data compared to QCD calculations with various nPDFs with and without inclusion of ${\rm D}^{0}$ measurements and FoCal pseudo-data.}
    \label{fig:directphotonRaa}
\end{figure}

Figure~\ref{fig:directphotonRaa} shows the nuclear modification factor $R_{\rm pPb}$ of inclusive direct photons at $\sqrt{s_{\rm NN}}=8.8\,{\rm TeV}$ compared to QCD calculations at NLO using nPDFs without constraints from ${\rm D}^{0}$ (gray), with re-weighting by ${\rm D}^{0}$ data (blue) or FoCal pseudo-data (red). The theoretical prediction without the re-weighting procedure (gray) has PDF uncertainties of approximately 30\% at $p_{\rm T}=5\,{\rm GeV}/c$. By applying the re-weighting procedure with FoCal pseudo-data, the PDF uncertainties are reduced by about 50\%, demonstrating the potential of FoCal measurements to constrain global PDF fits. The inclusion of ${\rm D}^{0}$ production at forward rapidity, measured by LHCb~\cite{D0LHCb}, results in a notable reduction of PDF uncertainties. Although this reduction is significant, exploring the low-$x$ phase-space requires a multi-messenger approach that includes a global analysis of all available data. Thus, incorporating FoCal direct photon data into global PDF fits will provide insights into factorisation and universality in nuclear exvironments.\\[2mm]
\emph{- Neutral meson measurements}

\begin{figure}[!hpt]
    \centering
    \includegraphics[width=0.48\textwidth]{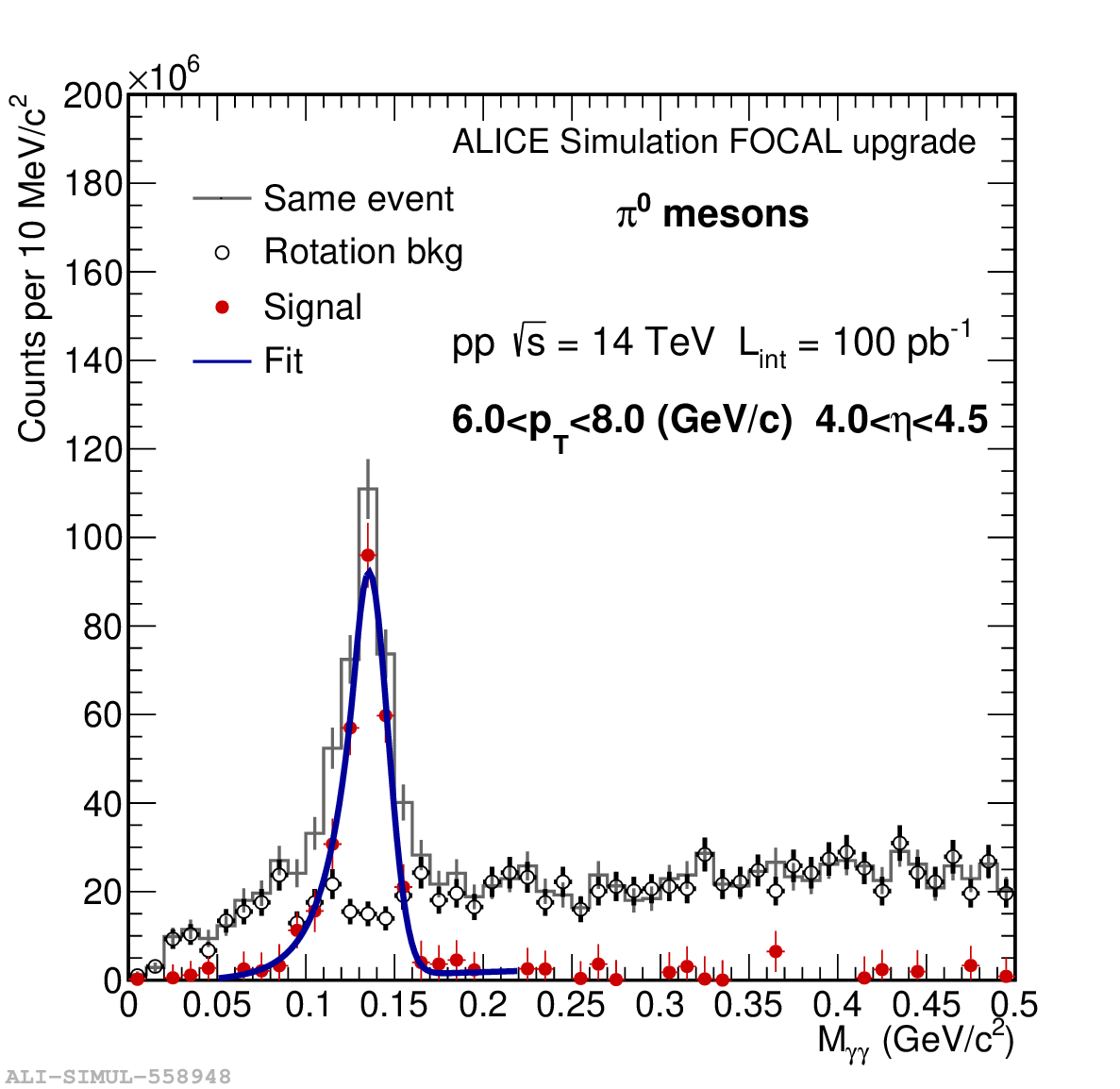}
    \caption{Invariant mass distribution of di-photons in pp collisions at $\sqrt{s}=14\,TeV$. The background distribution is determined by random rotation of the clusters from the same event.}
    \label{fig:pi0invmass}
\end{figure}

The FoCal can reconstruct neutral mesons (i.e. $\pi^{0}$, $\eta$, $\omega$, etc.) decaying into photons or electrons using the electromagnetic showers in FoCal-E. The reconstruction performance was studied using simulated pp collision events that include the underlying event, which contributes to the combinatorial background. Thus, the reconstructed invariant mass distribution includes several components: signal, cluster splitting, and combinatorial background. The distribution is fitted using a cocktail: the signal component from the distribution of cluster pairs matching true photon pairs from $\pi^{0}$ decays, the cluster-splitting component from an invariant mass analysis of single-photon events where no signal contribution is expected. Three different approaches were tested for handling the combinatorial background: a polynomial function fit, event mixing, and rotational background method. The first two methods showed similar performance in describing the combinatorial background, but they did not accurately capture the distribution below $\pi^{0}$ peak ($<0.1\,{\rm GeV}/c^{2}$) due to unaccounted correlated backgrounds, such as photon pairs from the splitting of a single $\pi^{0}$ or secondary clusters from other $\pi^{0}$ photons~\cite{focalphysicsperformance}. In contrast, the rotation method demonstrated better performance, as shown in Fig.~\ref{fig:pi0invmass}

Figure~\ref{fig:pi0invmass} shows an example of signal extraction using the rotation method, where the distribution is well described by both the signal component and rotational background, indicating that an additional template for cluster splitting is not necessary. After background subtraction, a clear signal distribution is fitted well by the Crystal Ball function. With the FoCal detector, the $\pi^{0}$ measurements will provide insights into gluon PDFs.\\[2mm]
\emph{- Jet measurements}

Forward jet measurements are sensitive to saturation effects at small-$x$, as there are three momentum scales involved in the process: $Q_{\rm sat}$, which characterises gluon-saturated matter at small-$x$, $p_{\rm T}^{\rm jet}$ of the individual jets, and the momentum imbalance $k_{\rm T}$ of the jet pair corresponding to the transverse momentum of the small-$x$ gluons involved in the hard scattering. The performance of jet reconstruction is assessed using the anti-$k_{\rm T}$ clustering algorithm~\cite{antikt} at both particle and detector levels. At the particle level, jet utilise particles within the FoCal acceptance, while at the detector level, jets use FoCal-E clusters and FoCal-H tower signals. Reconstruction performance is quantified by evaluating the relative energy and $p_{\rm T}$ difference between particle-level and detector-level jets, defined as

\begin{equation}
    \Delta E=\frac{E^{\rm det}-E^{\rm part}}{E^{\rm part}}, ~~~
    \Delta p_{\rm T}=\frac{p_{\rm T}^{\rm det}-p_{\rm T}^{\rm part}}{p_{\rm T}^{\rm part}}.\\\nonumber
\end{equation}

Two key metrics, Jet Energy Scale (JES) and Jet Energy Resolution (JER), are evaluated using the mean and RMS of the $\Delta E$ and $p_{\rm T}$ distributions.

\begin{figure}[!hpt]
    \centering
    \includegraphics[width=0.48\textwidth]{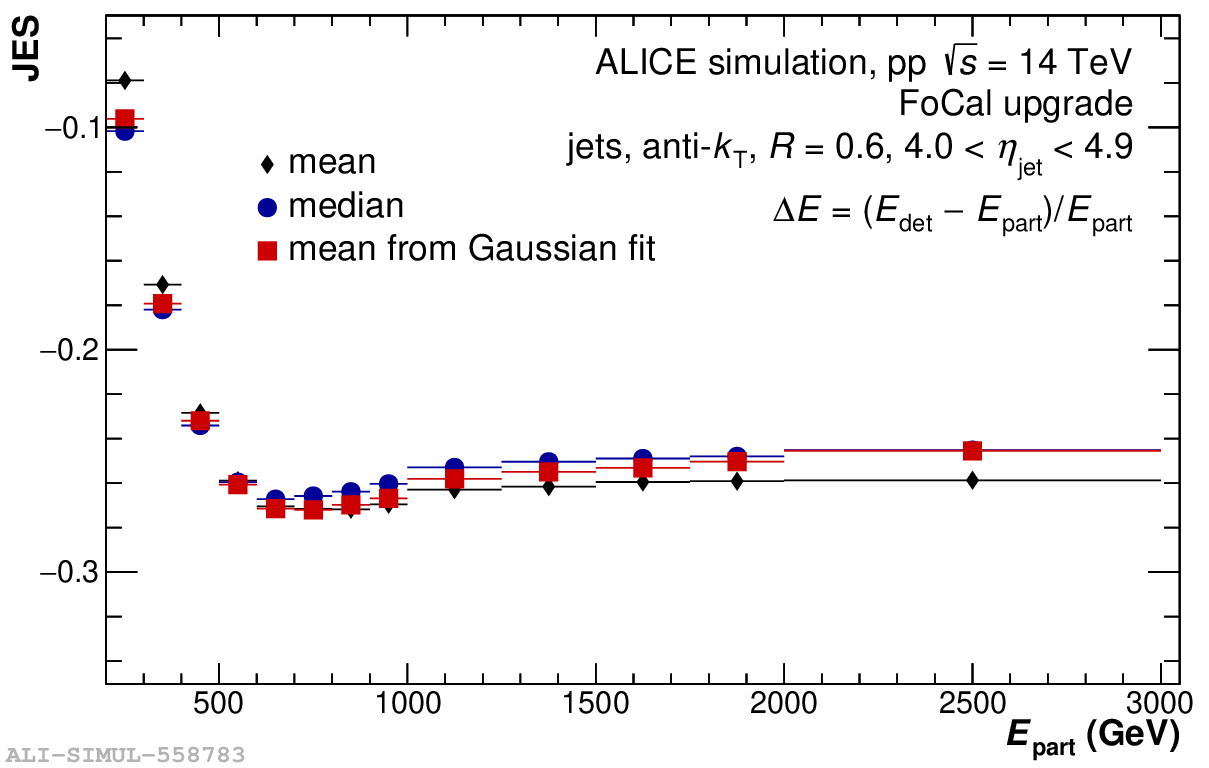}
    \caption{Jet Energy Scale for jets reconstructed using anti-$k_{\rm T}$ algorithm with $R=0.6$ in pp collisions at $\sqrt{s}=14\,{\rm TeV}$. The value is calculated using $\Delta E$.}
    \label{fig:jes}
\end{figure}

\begin{figure}[!hpt]
    \centering
    \includegraphics[width=0.48\textwidth]{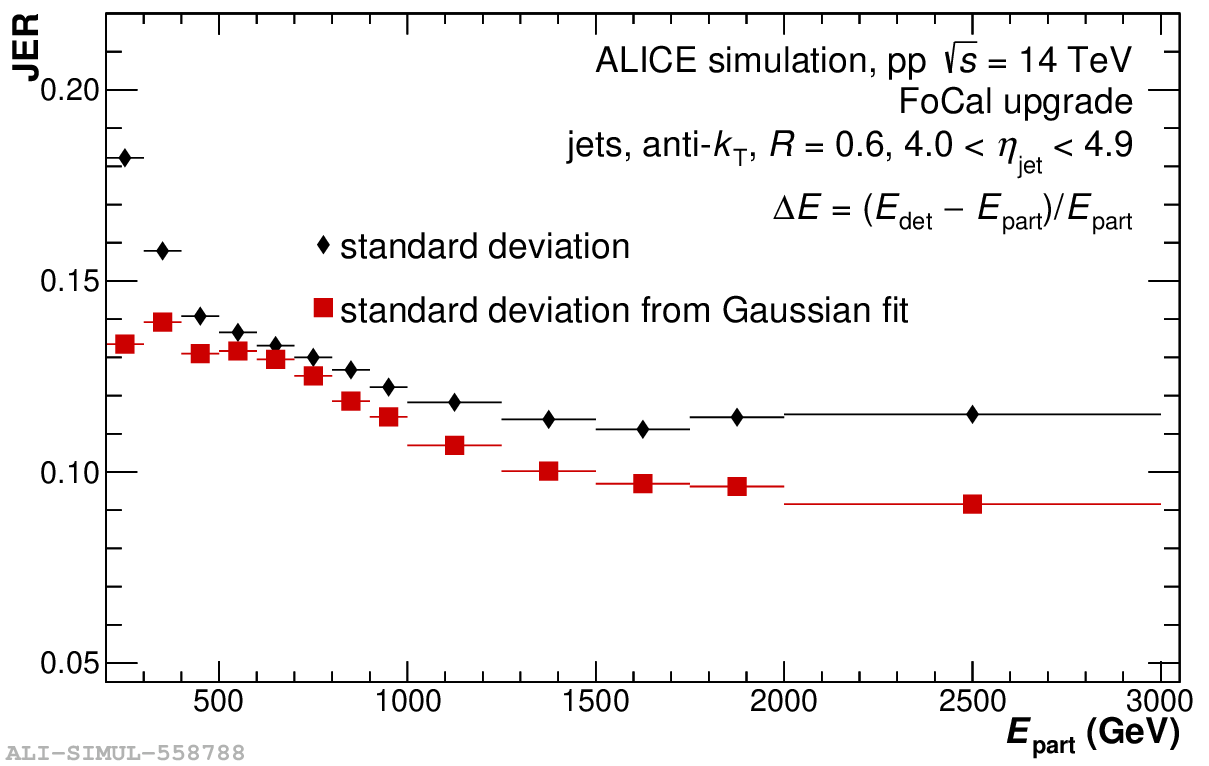}
    \caption{Jet Energy Resolution for jets reconstructed using anti-$k_{\rm T}$ algorithm with $R=0.6$ in pp collisions at $\sqrt{s}=14\,{\rm TeV}$. The value is calculated using $\Delta E$.}
    \label{fig:jer}
\end{figure}

The JES and JER for jets with $R=0.6$ and centroid $4.0<\eta^{\rm jet}<4.9$, calculated using $\Delta E$ as a function of the particle-level energy of the reconstructed jet, are shown in Fig.~\ref{fig:jes} and Fig.~\ref{fig:jer}, respectively. A negative JES value indicates that a fraction of the total jet energy escapes the jet reconstruction, resulting in a deficit in jet energy. The JES value decreases rapidly up to $600\,{\rm GeV}$ and remains relatively stable above this energy range. The JER is determined from Gaussian fits and numerical integration of the $\Delta E$ distributions separately. It remains below 15\% for energies up to $3\,{\rm TeV}$, although the value from the numerical integration exceeds 15\% below $400\,{\rm GeV}$. Further studies on the jets have been conducted, revealing potential improvements by accounting for biases in the neutral energy fraction of the jet population~\cite{focalphysicsperformance}.\\[2mm]
\emph{- Photon--hadron correlations}

Direct photon-hadron ($\gamma_{\rm dir}$--h) correlations in the forward region of pA collisions are sensitive to small-$x$ gluon dynamics. The gluon dynamics at small-$x$ are expected to modify the shape of the azimuthal distribution of $\gamma_{\rm dir}$--h correlations. The $\gamma_{\rm dir}$--h correlations can be determined from $\gamma_{\rm iso}$--h correlation by suppressing the non-$\gamma_{\rm dir}$--h correlation components. Thus, we performed a FoCal physics performance study for this channel by measuring the accuracy of the widths of $\gamma_{\rm iso}$--h correlation functions.

\begin{figure}[!hpt]
    \centering
    \includegraphics[width=0.48\textwidth]{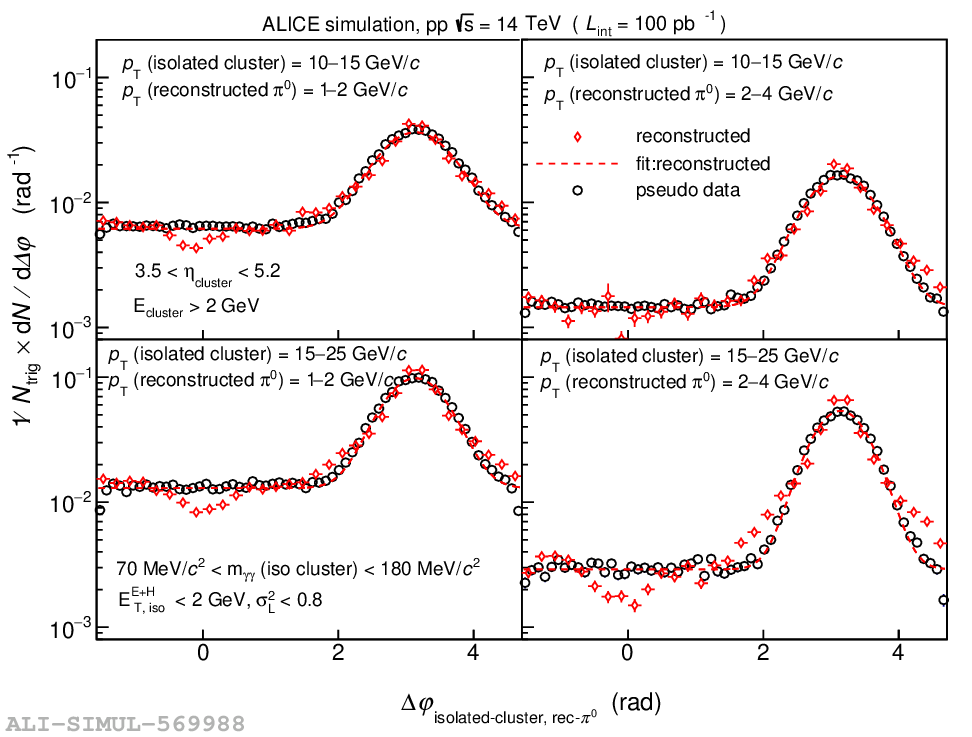}
    \caption{Azimuthal distribution of isolated cluster--$\pi^{0}$ correlation function in the FoCal acceptance in pp collisions at $\sqrt{s}$=14\,{\rm TeV}.}
    \label{fig:corr}
\end{figure}

Figure~\ref{fig:corr} shows the raw isolated cluster--$\pi^{0}$ candidate correlation function at the detector level (red marker) and the pseudo-data for the raw correlation function (black marker) in pp collisions at $\sqrt{s}=14\,{\rm TeV}$. The pseudo-data are obtained from a fit of the raw isolated cluster--$\pi^{0}$ correlation function using a Gaussian function. In the raw correlation function, there is no peak in the near-side region, but rather a small dip around $\Delta\varphi=0$ due to the isolation cut. On the other hand, no dip behaviour appears in the pseudo-data since the fit does not account for the behaviour at $\Delta\varphi=0$. The fit function also does not describe the simulation for the highest trigger and associated $p_{\rm T}$ bin, resulting in discrepancies between the simulation and pseudo-data as shown in the bottom-right panel of Fig.~\ref{fig:corr}. The pseudo-data are refit using the aforementioned function, and the width of the correlation function and its uncertainties are extracted from the fit.

\begin{figure}[!hpt]
    \centering
    \includegraphics[width=0.48\textwidth]{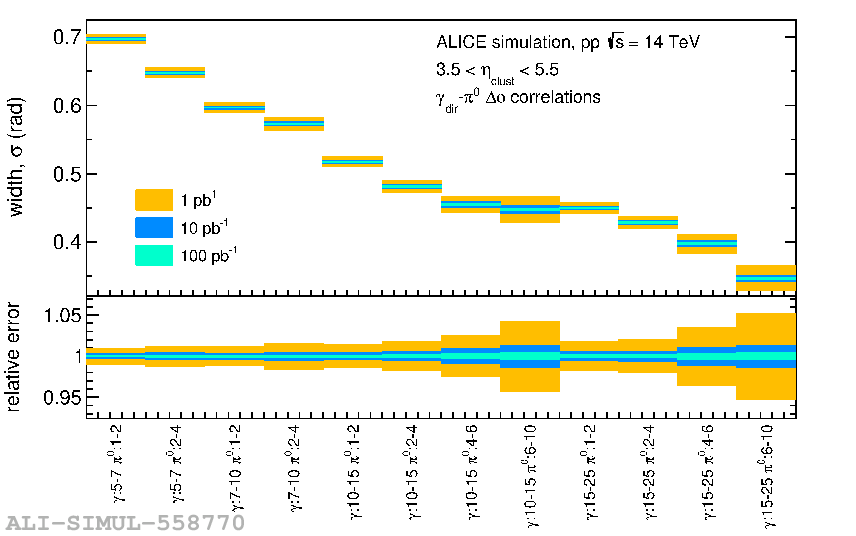}
    \caption{Extracted width and corresponding uncertainties from a fit to $\gamma_{\rm iso}$--$\pi^{0}$ correlation functions for given trigger and associated $p_{\rm T}$ in pp collisions at $\sqrt{s}=14\,{\rm TeV}$.}
    \label{fig:width}
\end{figure}

Figure~\ref{fig:width} shows the width of the correlation function from the fit and its corresponding uncertainties in selected trigger and associated $p_{\rm T}$ bins. The distributions become narrow for higher trigger and associated $p_{\rm T}$ bins, reflecting the collimated recoil jet peak. According to the simulation study, the relative statistical error is less than 1\% for the projected Run 4 integrated luminosity, $\mathscr{L}_{\rm int}=100\,{\rm pb}^{-1}$.\\[2mm]
\emph{- Vector meson photoproduction in ultra-peripheral collisions}

The photoproduction in ultra-peripheral collisions (UPC) extends the kinematic reach of current ALICE measurements and complement the EIC program~\cite{eic}. The photoproduction cross sections of heavy vector mesons are sensitive to gluon dynamics as per LO pQCD calculations. The FoCal provides unique kinematic coverage, significantly enhancing measurements of J/$\psi$ and $\psi(2S)$ photoproduction cross sections in p--Pb collisions at center-of-mass energies $W_{\gamma p}=2\,{\rm TeV}$ of the emitted photon and the proton projectile, extending up to $2\,{\rm TeV}$.

\begin{figure}[!hpt]
    \centering
    \includegraphics[width=0.48\textwidth]{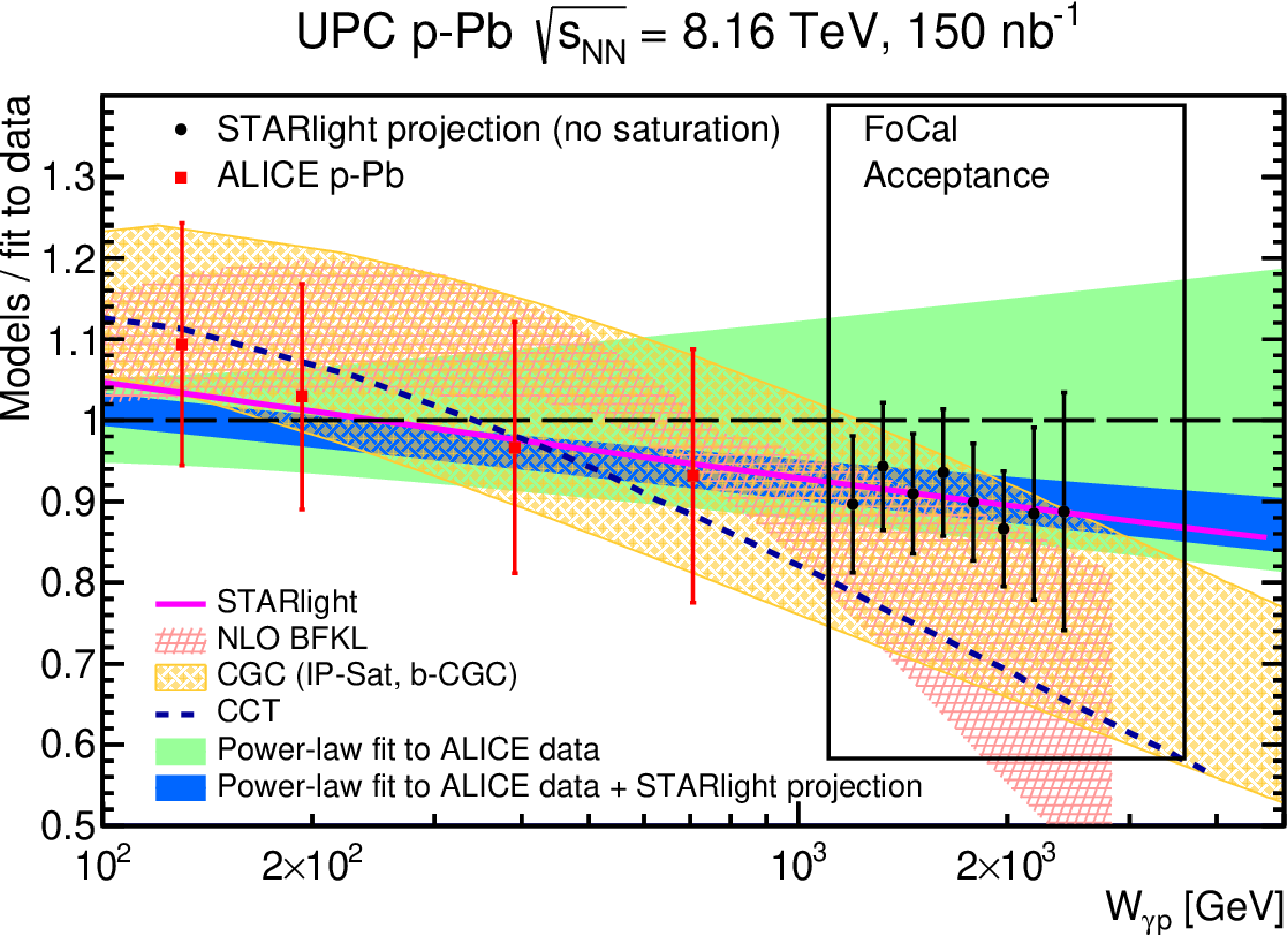}
    \caption{Ratio of the ALICE data and simulations to the power-law fit used for the ALICE data. Figure from \cite{upc}.}
    \label{fig:upc}
\end{figure}

\begin{figure}[!hpt]
    \centering
    \includegraphics[width=0.48\textwidth]{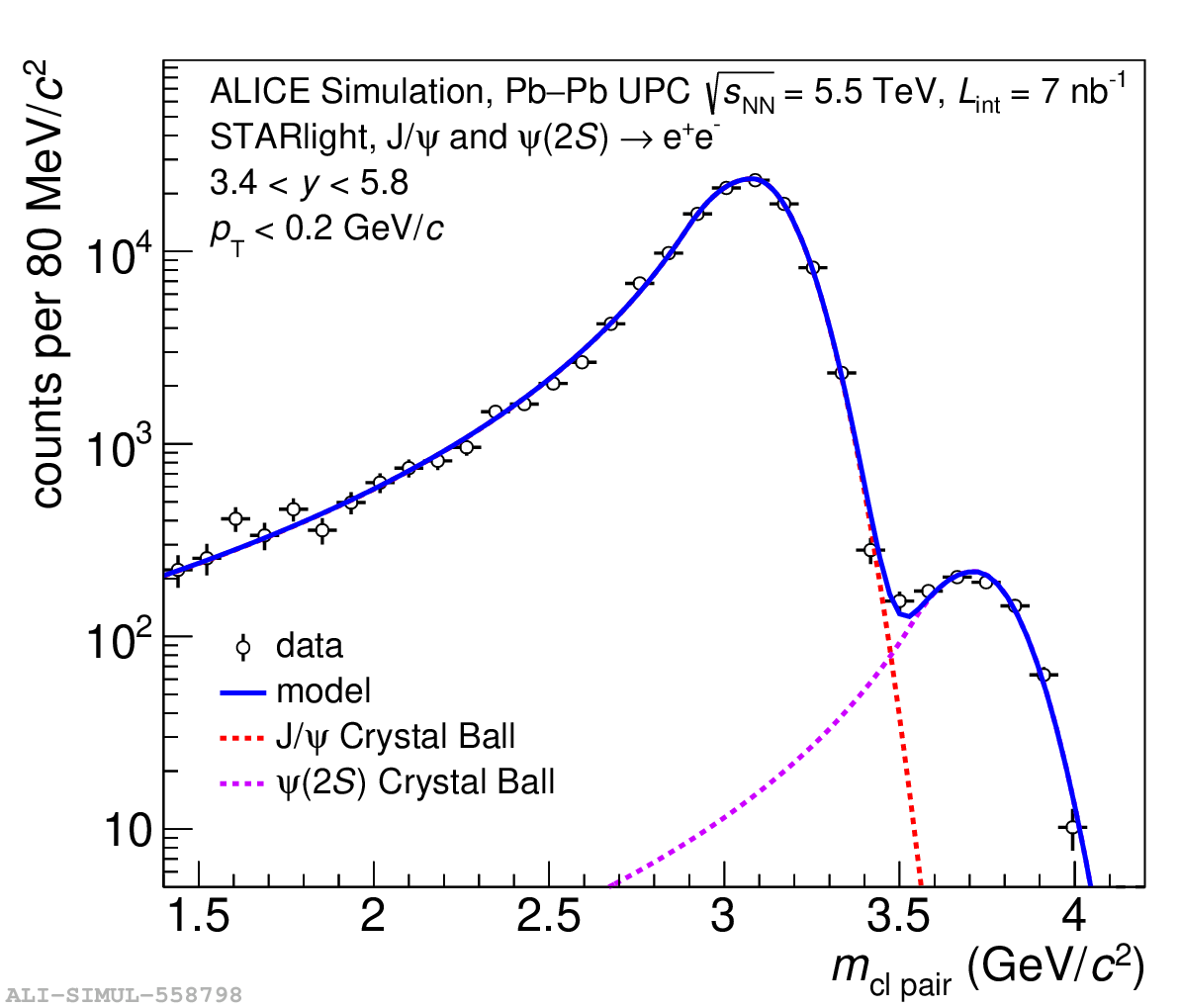}
    \caption{Invariant mass distribution for J/$\psi$ and $\psi(2S)$ in the FoCal acceptance.}
    \label{fig:invmass}
\end{figure}

Figure~\ref{fig:upc} shows the ratio of the ALICE data and NLO BFKL projection to the power-law used to fit the ALICE data. It demonstrates how FoCal's UPC measurements will reveal deviations from power-law growth at high energies, particularly when saturation occurs. Figure~\ref{fig:invmass} displays the invariant mass distribution for cluster pairs obtained from coherent J/$\psi$ and $\psi(2S)$ STARlight simulations~\cite{starlight}. The signal is extracted using a sum of double-sided Crystal Ball functions, clearly distingushing between the J/$\psi$ and $\psi(2S)$ states in FoCal. This illustrates successful measurements of J/$\psi$ and $\psi(2S)$ in ultra-peripheral Pb--Pb collisions.

\section{FoCal design concept}
As mentioned, FoCal consists of electromagnetic (FoCal-E) and hadronic (FoCal-H) components. FoCal-E is a Silicon(Si)+Tungsten(W) sampling calorimeter with fine lateral granularity readout. The W absorber material has a small Molière radius $R_{\rm M}\approx0.9\,{\rm cm}$ and radiation length $X_{\rm 0}=3.5\,{\rm mm}$, resulting in a total radiation length of approximately $20X_{0}$. FoCal-E comprises 18 silicon pad layers and two silicon pixel layers. The pad sensor has a transverse cell size of $1\,{\rm cm}^{2}$, which are read out by the High Granularity Calorimeter ReadOut Chip (HGCROC)~\cite{hgcroc}, allowing individual readout for channels in each layer. The HGCROC provides Analog to Digital Converter (ADC), Time Of Arrival (AOT) and Time Over Threshold (TOT). The ADC samples at 40\,MHz with a configurable phase shift to match the LHC bunch collision timing. TOA measures signal arrival time relative to the interaction, facilitating TOT computation, which extends the dynamic range of the ADC. The pixel sensor, with a pixel size of approximately $30\times30\,{\rm \mu m}^{2}$, is the ALICE Pixel Detector (ALPIDE)~\cite{alpide}, based on Monolithic Active Pixel Sensor (MAPS) technology. These sensors are located at the 5th and 10th layer. The readout chain for the pixel layers is similar to ITS2, but with modifications: the ALPIDEs are bonded in multi-chip strings using the SpTAP (Single-point Tape Automated Bonding) technique~\cite{focaltdr}, instead of wire bonding used in ITS2. The readout rates for the pixel layers are $1.2\,{\rm Gbps}$ and $400\,{\rm Mbps}$ for the Inner Barrel (IB) and Outer Barrel (OB), respectively. The FoCal-E module is followed by FoCal-H, a conventional hadronic sampling calorimeter with an effective nuclear interaction length of approximately $\sim5\lambda_{\rm int}$. FoCal-H lacks longitudinal segmentation and will be constructed using $2.5\,{\rm mm}$ outer diameter Cu capillary tubes filled with plastic scintillating fibers. The readout for FoCal-H utilise the H2GCROC (HGCROC for silicon photomultipliers(SiPMs)) in conjunction with SiPMs. Most functionalities of the H2GCROC are similar to those of the HGCROC used for the FoCal-E pads. The main difference lies in the use of a current conveyor in the analog input to attenuate the signal with a programmable gain. Additionally, the chip can adjust the bias voltage individually for each channel, tailoring it to the specific SiPMs to ensure a more uniform response.

\section{Performance of the FoCal prototype}
The performance of a full-length prototype of the FoCal detector has been studied through extensive test beam experiments at CERN PS and SPS from 2021 to 2023. The data were collected using hadron beams with energies up to 350\,GeV and electron beams with energies up to 300\,GeV. The performance of FoCal-E was studied by analysing the pad response to minimum ionising particles and quantifying the transverse shower width for shower separation in FoCal-E pixels. The performance of FoCal-H is studied by analysing the response to hadron beams and by comparing the data to simulation.



\begin{figure}[!hpt]
    \centering
    \includegraphics[width=0.45\textwidth]{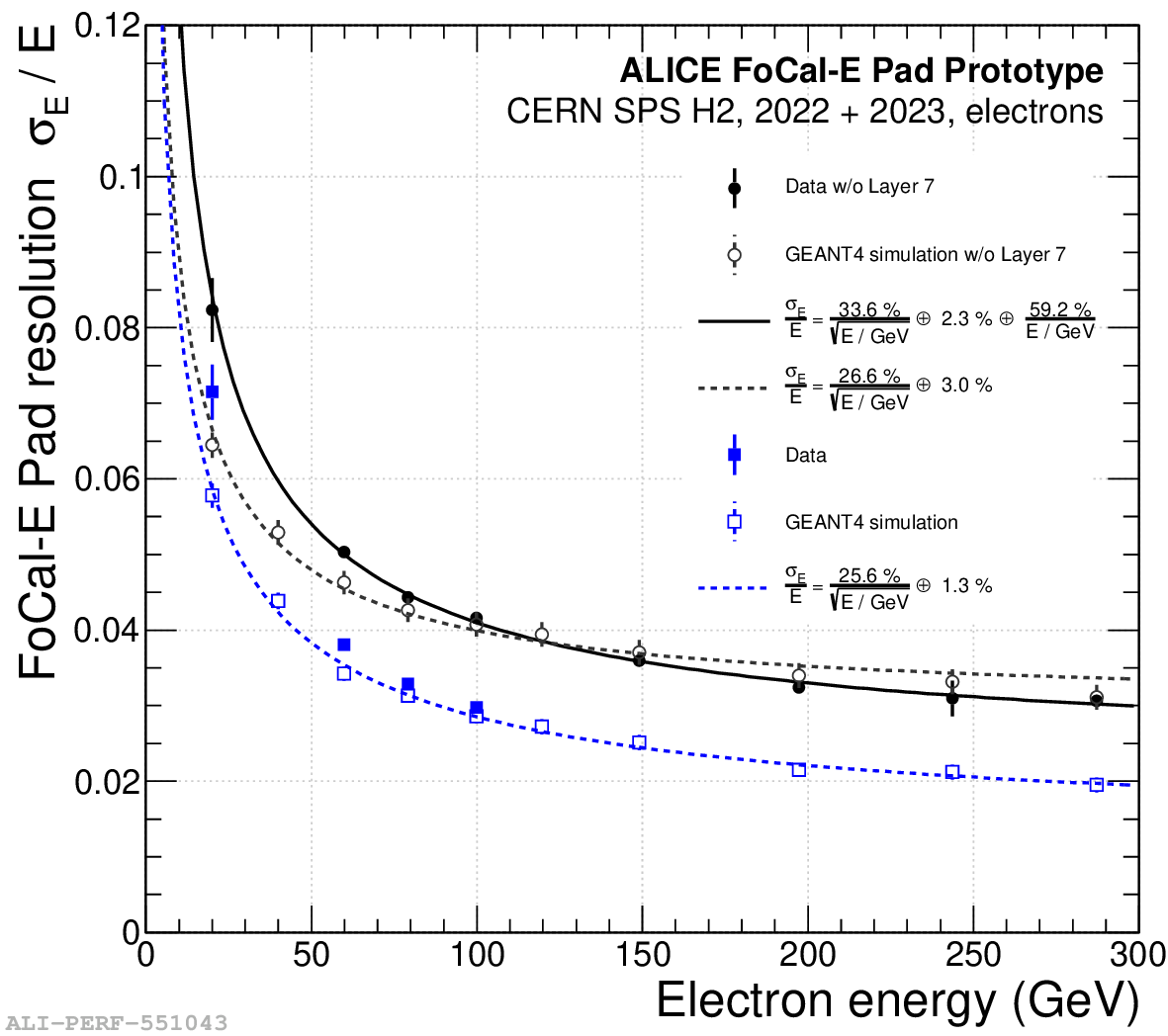}
    \caption{Relative energy resolution for the FoCal-E pad layers, compared to simulations, and respective fits.}
    \label{fig:epadtb}
\end{figure}

To evaluate the performance of the FoCal-E pads, the linearity and resolution were assessed using electron beams. The energy response to electrons was calibrated by summing the charge signals from all active pad layers. The distributions of summed pad signals were fitted with Gaussian curves, and the mean and width of these distributions were extracted from the fit parameters. The mean value as a function of electron energy was described by a linear fit, $Q(E)=q\times E+Q_{0}$, with the data showing less than 5\% deviation from the fit~\cite{focaltb}. The relative resolution is defined as $r(E)=\sigma_{Q}(E)/(Q(E)-Q_{0})$ where $Q_{0}$ is obtained from the linear fit. Figure~\ref{fig:epadtb} presents the relative energy resolution of the FoCal-E pad layers, compared with simulations. Both experimental and simulated resolution values agree within uncertainties for energies above 80\,GeV, and below 3\% for 100\,GeV, meeting the physics requirement of approximately 5\%.

\begin{figure}[!hpt]
    \centering
    \includegraphics[width=0.45\textwidth]{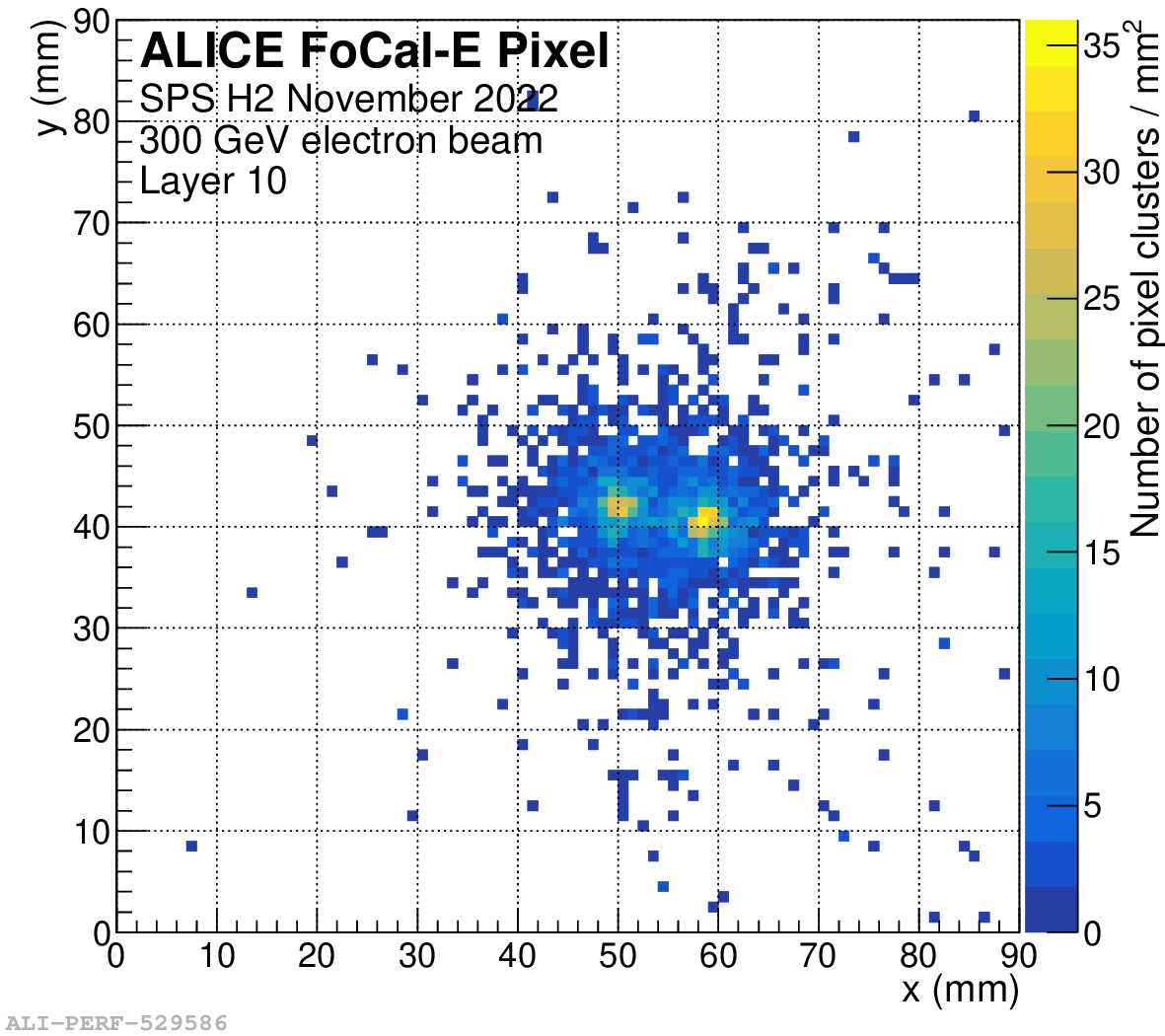}
    \caption{FoCal-E pixel event display of two-electron shower in layer 10.}
    \label{fig:epixeltb}
\end{figure}

The FoCal-E pixel layers enable shower separation, and their performance was evaluated using particle showers at sub-millimeter scales. As depicted in Fig.~\ref{fig:epixeltb}, the hit density distribution from a two-electron shower event with an energy of 300\,GeV exhibits clear separation over approximately 1\,cm. The distribution was projected onto a lateral axis, and the Full Width Half Maximum (FWHM) was measured to analyse the shower profile. The FWHM measures approximately 1.2\.mm (2.4\,mm) for an electron energy of 20\,GeV, reducing to 0.8\,mm (1.2\,mm) for an electron energy of 300\,GeV in layer 5 (layer 10)~\cite{focaltb}. Comparison with simulations show the measured data and simulations are within 0.5\,mm of each other.

\begin{figure}[!hpt]
    \centering
    \includegraphics[width=0.48\textwidth]{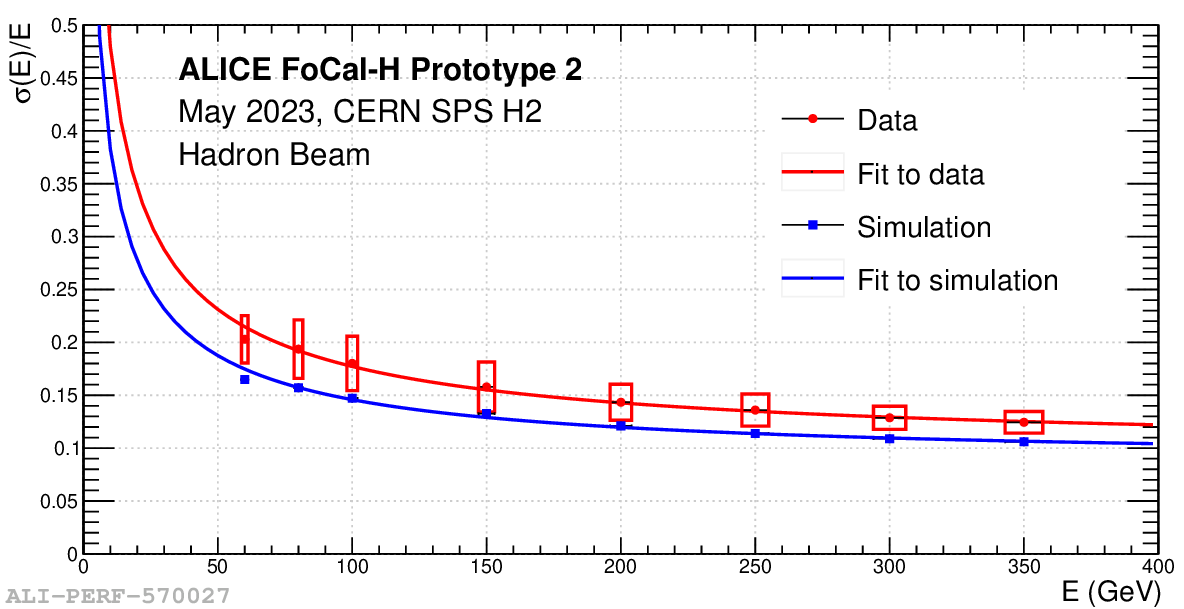}
    \caption{FoCal-H energy resolution as a function of hadron beam energies.}
    \label{fig:hcaltb}
\end{figure}

The performance of FoCal-H was assessed by analysing the ADC signal sum distribution with a beam energy scan up to 350\,GeV. The ADC distribution was fitted using a Gaussian curve to extract the mean and width, which characterise the detector response. Figure~\ref{fig:hcaltb} shows the energy resolution of FoCal-H as a function of beam energy. The resolution remains below 20\% across the entire energy range, with discrepancies from simulations less than 5\%.

\section{Summary and outlook}
FoCal aims to study gluon saturation. The capabilities of FoCal were verified through years of physics simulations and test beam experiments. These proceedings present projections of the performance for $\gamma_{\rm dir}$, $\pi^{0}$ mesons, jets, $\gamma$-h correlations, and J/$\psi$ in UPCs. Regarding detector performance, a comprehensive study of the detector prototypes was conducted using test beam experiments at CERN PS and SPS. FoCal is a project within ALICE, approved by the LHCC in March 2024, and is expected to be installed at the beginning of 2028. Several institutes involved in the project plan to proceed with mass production and module assembly from this year to mid-2027, and the detector installation is scheduled in the early 2028. Through commissioning in 2028, the FoCal will begin data-taking from 2029. The FoCal data will allow us to explore an unprecedented region of momentum fraction, down to $x\sim10^{-6}$.

\end{document}